\documentclass[traditabstract]{aa}
\usepackage{txfonts}
\usepackage{eepic}
\usepackage{float}
\usepackage{graphicx}
\usepackage{natbib}
\usepackage{bigstrut}
\usepackage{multirow}
\usepackage{subfigure}
\usepackage{amssymb}
\usepackage{color}
\usepackage[normalem]{ulem}
\newcommand{\WHz}{\mathrm{W\ Hz^{-1}}}

\newcommand{\angstrom}{\mathrm{\AA}}

\newcommand{\Msun}{\mathrm{M_{\odot}}} 
\newcommand{\Myr}{\mathrm{M_{\odot}/yr}} 
\newcommand{\Dn}{\mathrm{D_n4000}}

\begin{document}
\title{The triggering probability of radio-loud AGN}
\subtitle{A comparison of high and low excitation radio galaxies in hosts of different colors}
\author{R.M.J. Janssen \inst{1,2} \and H.J.A. R\"{o}ttgering \inst{1} \and P.N. Best \inst{3} \and J. Brinchmann \inst{1}}
\institute{Leiden Observatory, University of Leiden, PO Box 9513, 2300 RA Leiden, The Netherlands \\ \email{rottgering@strw.leidenuniv.nl} \and Kavli Institute of Nanoscience, Faculty of Applied Sciences, Delft University of Technology, Lorentzweg 1, 2628CJ Delft, The Netherlands \\ \email{r.m.j.janssen@tudelft.nl} \and SUPA, Institute for Astronomy, University of Edinburgh, Royal Observatory, Blackford Hill, Edinburgh EH9 3HJ, UK}
\date{Received 16 February 2012 / Accepted 21 March 2012}

\abstract{Low luminosity radio-loud active galactic nuclei (AGN) are generally found in massive red elliptical galaxies, where they are thought to be powered through gas accretion from their surrounding hot halos in a radiatively inefficient manner. These AGN are often referred to as ``low-excitation'' radio galaxies (LERGs). When radio-loud AGN are found in galaxies with a young stellar population and active star formation, they are usually high-power radiatively-efficient radio AGN (``high-excitation'', HERG). Using a sample of low-redshift radio galaxies identified within the Sloan Digital Sky Survey (SDSS), we determine the fraction of galaxies that host a radio-loud AGN, $f_{RL}$, as a function of host galaxy stellar mass, $M_*$, star formation rate, color (defined by the 4000 $\angstrom$ break strength), radio luminosity and excitation state (HERG/LERG). \\We find the following: 1. LERGs are predominantly found in red galaxies. 2. The radio-loud AGN fraction of LERGs hosted by galaxies of any color follows a $f^{LE}_{RL} \propto M^{2.5}_*$ power law. 3. The fraction of red galaxies hosting a LERG decreases strongly for increasing radio luminosity. For massive blue galaxies this is not the case. 4. The fraction of green galaxies hosting a LERG is lower than that of either red or blue galaxies, at all radio luminosities. 5. The radio-loud AGN fraction of HERGs hosted by galaxies of any color follows a $f^{HE}_{RL} \propto M^{1.5}_*$ power law. 6. HERGs have a strong preference to be hosted by green or blue galaxies. 7. The fraction of galaxies hosting a HERG shows only a weak dependence on radio luminosity cut. 8. For both HERGs and LERGs, the hosting probability of blue galaxies shows a strong dependence on star formation rate. This is not observed in galaxies of a different color.\\Our interpretation of these results is that the presence of cold gas in a LERG enhances the probability that its SMBH becomes a luminous radio-loud AGN compared to the typical ``model'' LERG in a red elliptical galaxy. If enough cold gas can be transported to the SMBH a HERG can be created. However, the presence of cold gas does not automatically imply a HERG will be created. We speculate that feedback of the enhanced AGN activity in blue galaxies is responsible for the reduced probability of green galaxies to host a LERG.}
\maketitle

\section{Introduction}
It has been known for a long time that low luminosity radio-loud (RL) active galactic nuclei (AGN) are predominantly hosted by massive red elliptical galaxies \citep[e.g.][]{Matthews1964}. These AGN are thought to be the result of an accretion process in which hot gas from the red elliptical's X-ray halo is falling at low Eddington rates into the super-massive black hole (SMBH) at its center \citep[e.g.][]{Best2006}. Because this accretion process is radiatively inefficient, halo-fed radio galaxies often lack the luminous emission lines, X-ray emission and infrared torus emission generally associated with AGN \citep[e.g.][]{Laing1994,Evans2006,Whysong2004}. Therefore, they have been referred to as ``radio-mode'' or ``low-excitation'' radio galaxies (LERGs). A limited number of LERGs have also been found in massive galaxies with a younger stellar population \citep[bluer color; e.g.][]{Tadhunter2005}. Because young stars require cold gas to form, a young stellar population is often associated with the presence of a cold gas reservoir in the galaxy. If gas from this reservoir reaches the SMBH, it can be accreted at higher rates in a radiatively efficient process, creating an AGN with many luminous emission lines. This accretion process is more similar to the ``standard'' AGN accretion mode found in quasars. Hence, the population of emission line rich RL AGN is referred to as ``quasar-mode'' or ``high-excitation'' radio galaxies (HERGs).\\
High and low excitation radio galaxies were shown to be two independent populations \citep{Best2005b,Tasse2008b,Hickox2009} in which HERGs are typically hosted by lower mass, bluer galaxies \citep{Best2012} located in less dense environments. Despite this preference, LERGs have been identified in blue galaxies and HERGs in red galaxies. Because these populations are rare, the number of observed examples has not been large enough for a statistical study. However, using the Sloan Digital Sky Survey \citep[SDSS; ][and references therein]{York2000,Stoughton2002} in combination with all sky radio surveys, a sample of radio galaxies can now be constructed that contains the number statistics to analyze the host galaxies of RL AGN in detail. Hereby we aim to achieve new insights into the effect of the presence or absence of an internal reservoir of cold gas on the creation of a RL AGN.\\
The sample of galaxies and RL AGN used to investigate this is described in Sect. \ref{Sample}. After defining the RL AGN fraction in Sect. \ref{FractionDef}, we use this parameter in Sect. \ref{Excitation} and \ref{Color} to investigate the probability of hosting a RL AGN as a function of host galaxy mass, SFR, color and the RL AGN's radio luminosity and excitation. An interpretation of these results is discussed in Sect. \ref{Discussion} before we summarize our conclusions in Sect. \ref{Conclusions}. Throughout this paper we assume the cosmological parameters $H_{0} = 70$ $\mathrm{km}$ $\mathrm{s^{-1}}$ $\mathrm{Mpc^{-1}}$, $\Omega_{m}=0.3$, and $\Omega_{\Lambda}=0.7$.
\section{\label{Sample}The Radio-loud AGN Sample}
\begin{figure}
\begin{center}
\includegraphics[%
  width=0.90\linewidth,
  keepaspectratio]{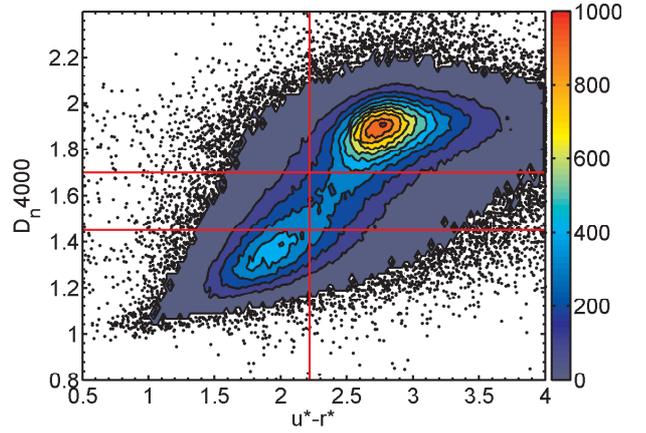}
\end{center}
\caption{A contour plot showing the distribution of galaxies with a mass $10^{10.25} \ \Msun \leq M_* \leq 10^{12.0} \ \Msun$ within the $u^*-r^*$ vs $\Dn$ plane. The colorbar on the right indicates the number of galaxies per bin of width 0.025 in both $u^*-r^*$ and $\Dn$. If the number of galaxies per bin is less than 5, these galaxies are shown as black dots instead. The vertical red line indicates the optimum separation in the Petrosian color, $u^*-r^*=2.22$, proposed by \citet{Strateva2001}. Although this does indeed separate galaxies of the blue cloud and red sequence, the separation between the two populations is more pronounced in $\Dn$. Therefore the horizontal separation lines $\Dn = 1.45$ and $\Dn = 1.7$ are adopted in this research. Galaxies between these two values are classified as green.}
\label{FigColorURD4}
\end{figure}

In this research we use the sample of RL AGN defined by \citet{Best2012}. It was constructed by cross-comparing the positions of galaxies in the seventh data release \citep[DR7,][]{Abazajian2009} of the SDSS spectroscopic sample with data from the National Radio Astronomy Observatory (NRAO) Very Large Array (VLA) Sky Survey \citep[NVSS,][]{Condon1998} and the Faint Images of the Radio Sky at Twenty centimeters survey \citep[FIRST,][]{Becker1995}. The optical parent sample consists of all galaxies in the value-added spectroscopic catalogs (VASC) created by the Max Planck Institute for Astrophysics and Johns Hopkins University group \citep[cf.][]{Brinchmann2004b}. These objects are either targets from the SDSS main galaxy sample \citep{Strauss2002} with a usable spectrum, or QSOs with a spectrum that is clearly dominated by the emission of stars in the host galaxy. For each galaxy the VASC contains a number of derived quantities such as the stellar mass, $M_*$, and star formation rate, SFR \citep{Kauffmann2003a,Kauffmann2003b}. Within the 927,522 galaxies of the VASC, \citet{Best2012} identified a magnitude-limited sample of $18286$ radio sources. This sample was shown to have a completeness of $95\%$ and a reliability of $99\%$ \citep{Best2005a}. Besides RL AGN, it also contains radio sources in which the radio emission is associated with star formation. \citet{Best2012} separated the RL AGN and star forming galaxies using three different criteria: emission line ratio diagnostics, such as the `BPT' diagram \citep{Baldwin1981}; the radio luminosity, $L_{1.4GHz}$, versus $H\alpha$ emission line luminosity \citep{Kauffmann2008}; the $L_{1.4GHz}/M_*$ vs $\Dn$ plane \citep{Best2005a}. Where possible, \citet{Best2012} also classified each true RL AGN as a ``low-excitation'' (9863 sources) or ``high-excitation'' (481 sources) radio galaxy, using emission line properties. We will use these classification in the research presented below.\\
From all the galaxies in the VASC we select the best observations of all unique objects with a reliable redshift in the range $0.03\leq z\leq 0.3$. This redshift range is narrow enough to exclude most evolutionary effects, whilst it is wide enough to keep the number statistics required for this work. We only consider galaxies in the mass range $10^{10.25} \ \Msun \leq M_* \leq 10^{12.0} \ \Msun$, because of the limited number of RL AGN available outside this mass range. Within the galaxy population a small contamination of stars and QSOs was found. These objects were removed based on their point source nature, because galaxy parameter estimation is unreliable for these sources. Visual inspection showed that our method removed all stars and QSOs.\\
A separation between galaxies of the red sequence, green valley and blue cloud is made using the restframe 4000 $\angstrom$ break value \citep[$\Dn$; ][]{Balogh1999}. In previous works this has typically been done using broad band colors. For example, \citet{Strateva2001} show that the optimal separation in the $u^*-r^*$ color based on Petrosian magnitudes is $u^*-r^*=2.22$. This separation is indicated by the vertical red line in Fig. \ref{FigColorURD4}, which shows the distribution of galaxies in the $u^*-r^*$ vs $\Dn$ plane.  However, from this figure it is clear that the galaxy color bimodality has a wider separation in $\Dn$ than $u^*-r^*$. Therefore we define galaxies with $\Dn \geq 1.7$ as ``red'' and galaxies with $\Dn \leq 1.45$ as ``blue''. Any galaxy with an intermediate value $(1.45<\Dn<1.7)$ is considered a ``green'' galaxy. These two separation values are shown as horizontal red lines in Fig. \ref{FigColorURD4}.\\
The number of galaxies (including those hosting a RL AGN) of each color can be found in Table \ref{GalaxyNumbers}. This table also gives the number of red, green and blue HERGs and LERGs for 2 different radio luminosity cuts ($L_{1.4GHz} \geq L_{1.4GHz}^{\min}$).\\
\begin{table}[b!]
\centering \caption{The number of red, green and blue galaxies and radio-loud AGN with a mass $10^{10.25} \ \Msun \leq M_* \leq 10^{12.0} \ \Msun$. The number of high and low excitation radio galaxies is given for 2 radio luminosity cuts, $L_{1.4GHz} \geq L_{1.4GHz}^{\min}$. Here $L_{1.4GHz}^{\min}$ is given in $\WHz$. Note that the number of galaxies includes the galaxies that host a radio-loud AGN.}
\begin{tabular}{cc|c|c|c|c|}
\cline{3-6}
 & & \multicolumn{2}{|c|}{LERG, $L_{1.4GHz}^{\min}$} & \multicolumn{2}{|c|}{HERG, $L_{1.4GHz}^{\min}$} \bigstrut \\
\hline
\multicolumn{1}{|c|}{Color} & \multicolumn{1}{|c|}{Galaxies} & $10^{23.0}$ & $10^{24.5}$ & $10^{23.0}$ & $10^{24.5}$ \bigstrut \\ 
\hline
\multicolumn{1}{|c|}{Red} & \multicolumn{1}{|c|}{295523} & 6157 & 1546 & 94 & 40 \\
\multicolumn{1}{|c|}{Green} & \multicolumn{1}{|c|}{114210} & 363 & 89 & 98 & 51 \\
\multicolumn{1}{|c|}{Blue} & \multicolumn{1}{|c|}{114947} & 171 & 83 & 60 & 40 \\
\hline
\end{tabular}
\label{GalaxyNumbers}
\end{table}
\section{\label{FractionDef}The Radio-loud AGN Fraction}
\begin{figure}
\begin{center}
\includegraphics[width=0.90\linewidth,keepaspectratio]{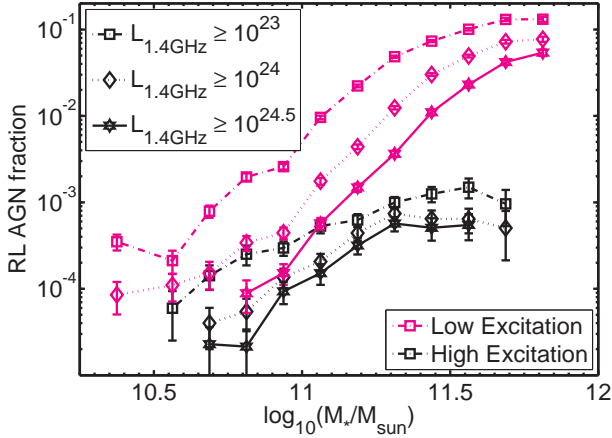}
\end{center}
\caption{The fraction of galaxies hosting a high (black) or low (magenta) excitation radio galaxy as a function of host galaxy stellar mass. Three radio luminosity cuts are given for each excitation. The errorbars are determined by Poissonian statistics.}
\label{FigFracERG}
\end{figure}
\begin{figure*}
	\begin{minipage}{0.48\textwidth}
	\centering
	\includegraphics[width=1\textwidth,keepaspectratio]{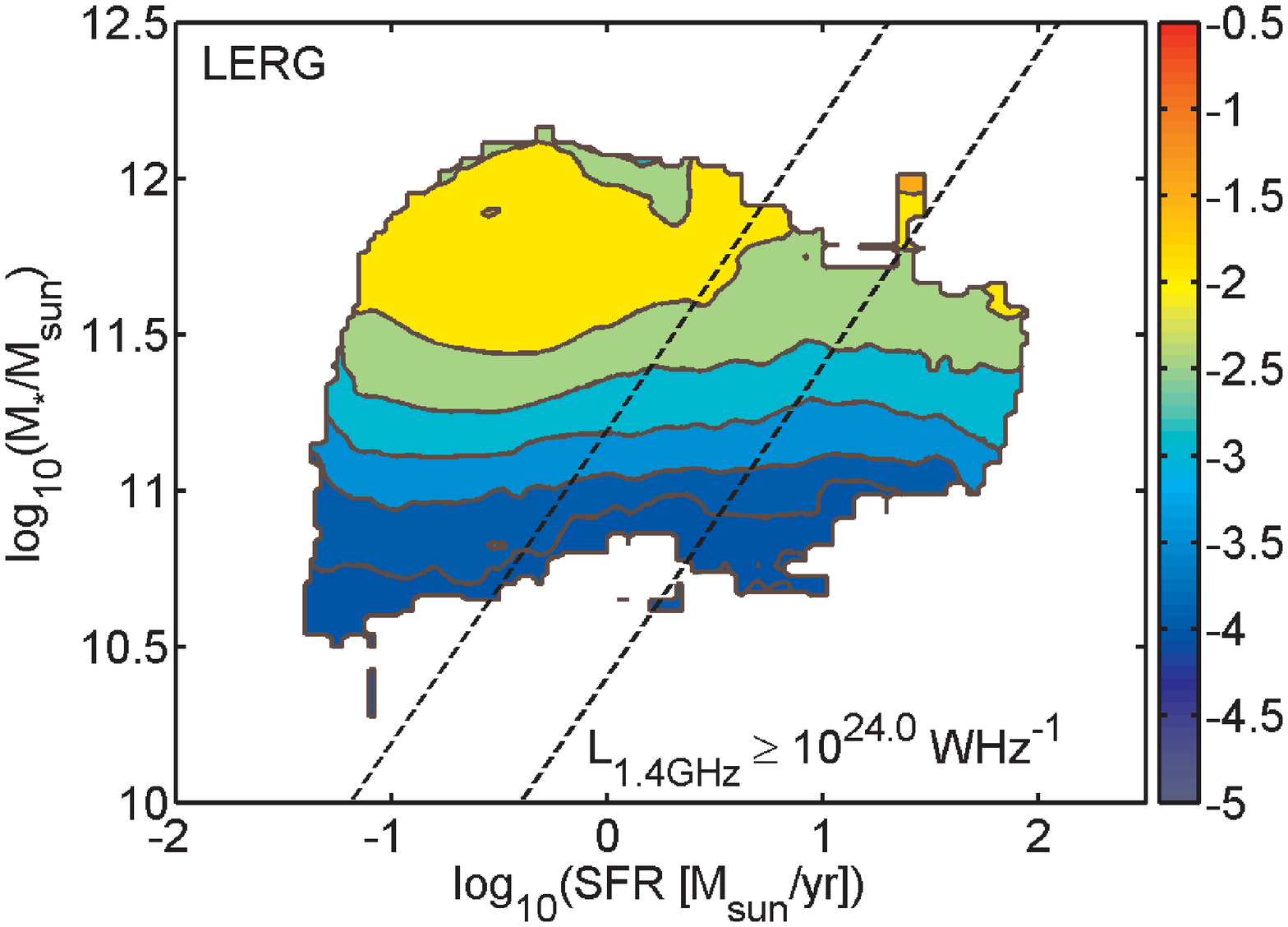}
	\end{minipage}
	\hspace{0.04\textwidth}
	\begin{minipage}{0.48\textwidth}
	\centering
	\includegraphics[width=1\textwidth,keepaspectratio]{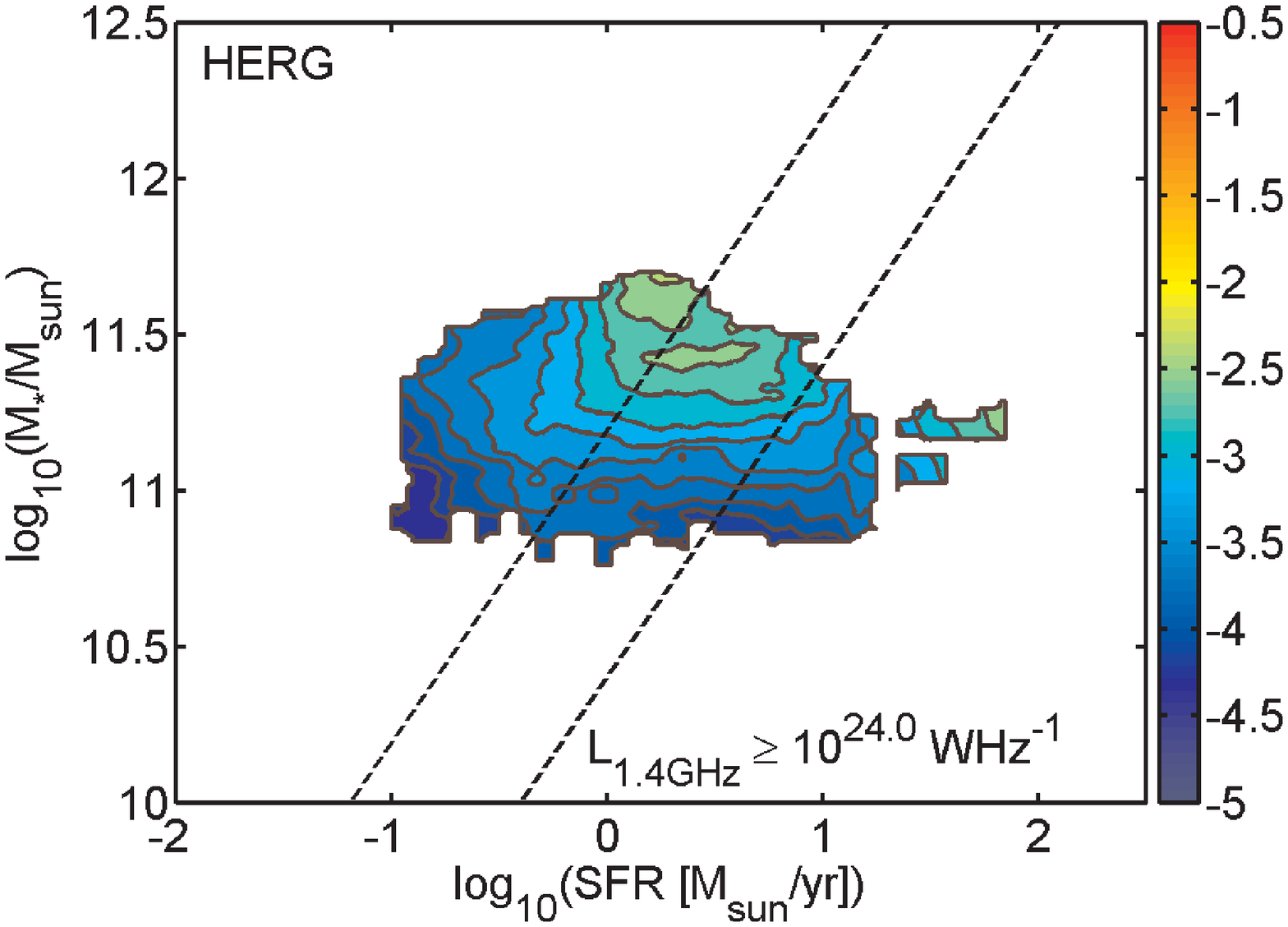}
	\end{minipage}
\caption{This contour plot shows the fraction of galaxies hosting a RL AGN in the stellar mass versus star formation rate plane. The value of $log_{10}(f^y_{RL})$ for low-excitation (left panel) and high-excitation (right panel) radio galaxies is given by color gradients for which the legend is found in the colorbars on the right. Only radio-loud AGN with a $L_{1.4GHz} \geq 10^{24.0}$ $\WHz$ were considered in the determination of $f^{y}_{RL}$. As the specific star formation rate $(\mathrm{SFR}/M_*)$ is broadly correlated with the $\Dn$ value of a galaxy, each panel contains 2 diagonals, which are a guide to the eye on where the red, green and blue galaxy population are located.}
\label{FigContourE}
\end{figure*}
We will investigate which galaxies preferentially host RL AGN by determining the fraction of RL AGN, $f_{RL}$, as a function of host galaxy mass, SFR, color ($x$) and the RL AGN's radio luminosity and excitation $(y)$. $f^{y,x}_{RL}$ is defined as
\begin{equation}
f^{y,x}_{RL} = \left(\sum_{i \in A_{y,x}}{\frac{1}{V_{\max}^i}}\right) \left(\sum_{j \in G_{x}}{\frac{1}{V_{\max}^j}} \right)^{-1}
\end{equation}
Here $A$ and $G$ are all RL AGN and galaxies (including AGN), respectively, in a particular bin. $V_{max}$ is the maximum volume in which each galaxy can be observed, as defined by the $\mathrm{V_{\max}}$-formalism \citep{Schmidt1968}. This takes into account the redshift limits from both the radio and optical selection criteria, as well as the limits due to the redshift range being analyzed. Previous works \citep[e.g. ][]{Best2005b} limited the volume studied at each radio luminosity to produce a volume limited sample. This allowed them to use simple fractions defined as:
\begin{equation}
f^{y,x}_{RL}=N^{y,x}_A/N^x_G
\end{equation}
Here $N_A$ and $N_G$ are the number of RL AGN and galaxies in a bin. Such analysis produces results that are equivalent to the results presented below, within the errors. We estimate the error on the RL AGN fraction, $\sigma(f_{RL})$, using Poissonian statistics \citep{Wall2003}.\\
\section{\label{Excitation} Dependence on AGN Excitation}
 Figure \ref{FigFracERG} shows the fraction of galaxies hosting either a HERG (black) or LERG (magenta) more luminous than a given radio power, as a function of host galaxy mass. At all radio luminosities the LERG population shows an increase in the RL AGN fraction that is consistent with $f^{LE}_{RL} \propto M^{2.5}_*$. The fraction of galaxies hosting a LERG saturates at high masses $(M_*>10^{11.6} \ \Msun)$ at a value of $\sim 10\%$. These results are similar to what was found by \citet{Best2005b}. Compared to the LERGs, HERGs have a far more shallow dependence on mass, $f^{HE}_{RL} \propto M^{1.5}_*$. In addition, the level of $f^{HE}_{RL}$ is much less dependent on the radio luminosity cut. Over the 1.5 orders of magnitude in radio luminosity shown in Fig. \ref{FigFracERG}, the fraction of galaxies hosting a HERG typically decreases by a factor of 3. This is significantly less than the factor of 10 decrease found at masses $M_*< 10^{11.5} \ \Msun$ between $f^{LE}_{RL}$ at the highest and lowest radio luminosity cuts.\\
As discussed in the introduction, the presence or absence of cold gas is thought to be of strong influence on the creation of either a HERG or LERG. Therefore it is of interest to consider the fraction of galaxies hosting a HERG or LERG as a function of SFR. Figure \ref{FigContourE} shows the SFR vs $M_*$ plane in which colored contours are used to indicate the value of $\log_{10}(f^y_{RL})$. The value of $\log_{10}(f^y_{RL})$ corresponding to each color can be found in the colorbars on the right of each panel. In the determination of $f^{y}_{RL}$ only RL AGN with a $L_{1.4GHz} \geq 10^{24.0}$ $\WHz$ are considered.\\
The ``colorbands'' of the LERG population are all horizontally aligned over the considered range of SFR, as can be seen in the left-hand panel of Fig. \ref{FigContourE}. Only for galaxies with both a high mass $(M_* \geq 10^{11.5} \ \Msun)$ and a high SFR ($\mathrm{SFR} \geq 10 \ \Myr$) is the SFR independence lost. However, given that there is little data in this region, this could very well be due to the scarceness of high mass, high SFR galaxies in general. The right-hand panel in Fig. \ref{FigContourE} shows the probability to host a HERG. From this figure it is clear that HERGs are preferentially hosted by a population of massive, $M_* \approx 10^{11.5} \ \Msun$, galaxies with a typical $\mathrm{SFR} = 3 \ \Myr$. From this maximum the hosting probability rapidly decreases to both lower mass and lower SFR.\\
The correlation between specific star formation rate $(\mathrm{sSFR} = \mathrm{SFR}/M_*$) and $\Dn$ \citep{Brinchmann2004b} allows us to give a guide to the eye on where the red, green and blue galaxy population approximately lie in the $(\mathrm{SFR},M_*)$-plane. The left and right dashed diagonal lines roughly correspond to $\Dn=1.7$ and $\Dn=1.45$, respectively. However, there is considerable overlap between galaxies of the three colors. Therefore, we will investigate the properties of the HERG and LERG population in red, green and blue galaxies individually in the next section.\\
\section{\label{Color}Dependence on Galaxy Color}
\begin{figure*}
	\begin{minipage}{0.48\textwidth}
	\centering
	\includegraphics[width=1\textwidth,keepaspectratio]{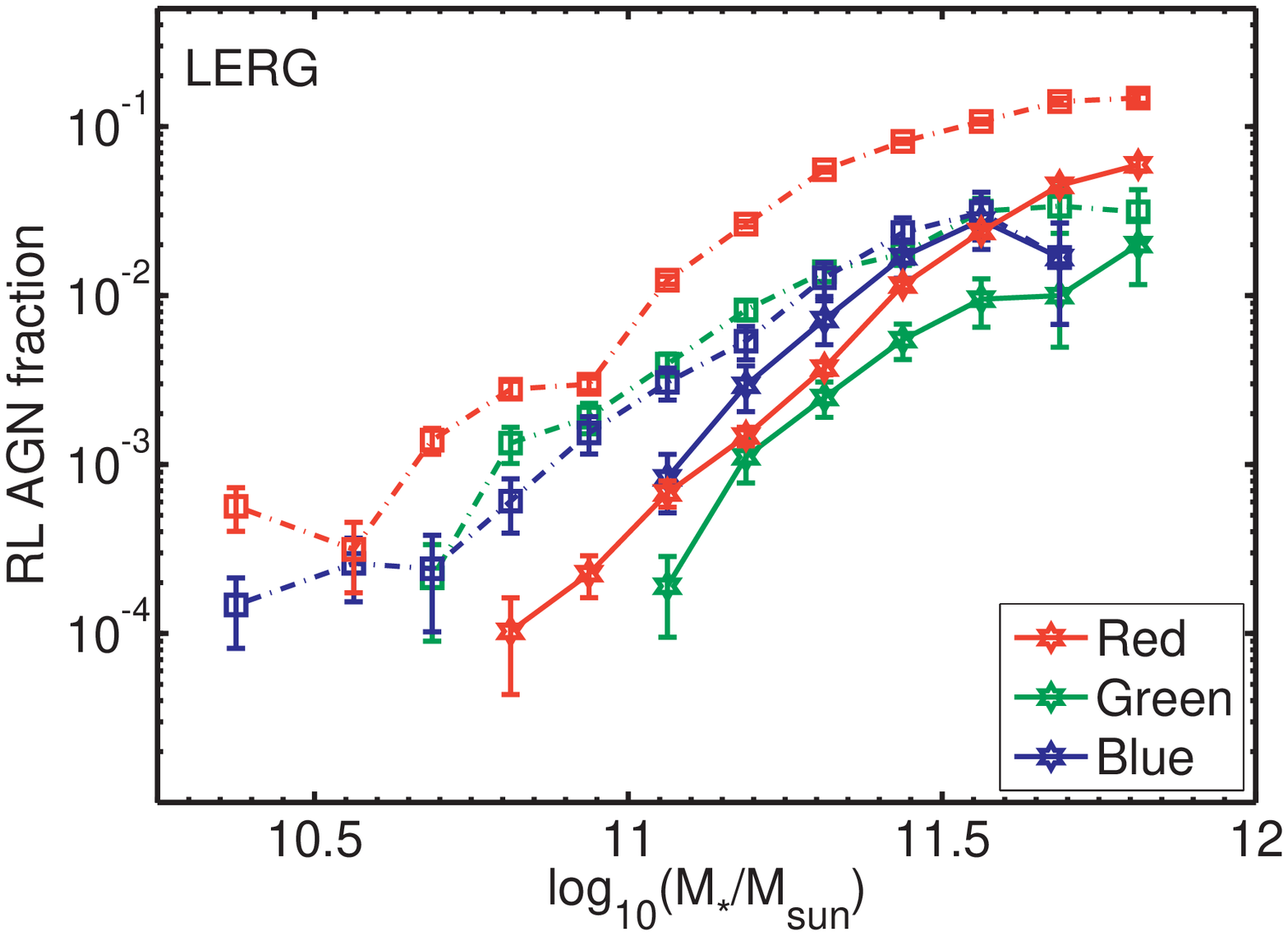}
	\end{minipage}
	\hspace{0.04\textwidth}
	\begin{minipage}{0.48\textwidth}
	\centering
	\includegraphics[width=1\textwidth,keepaspectratio]{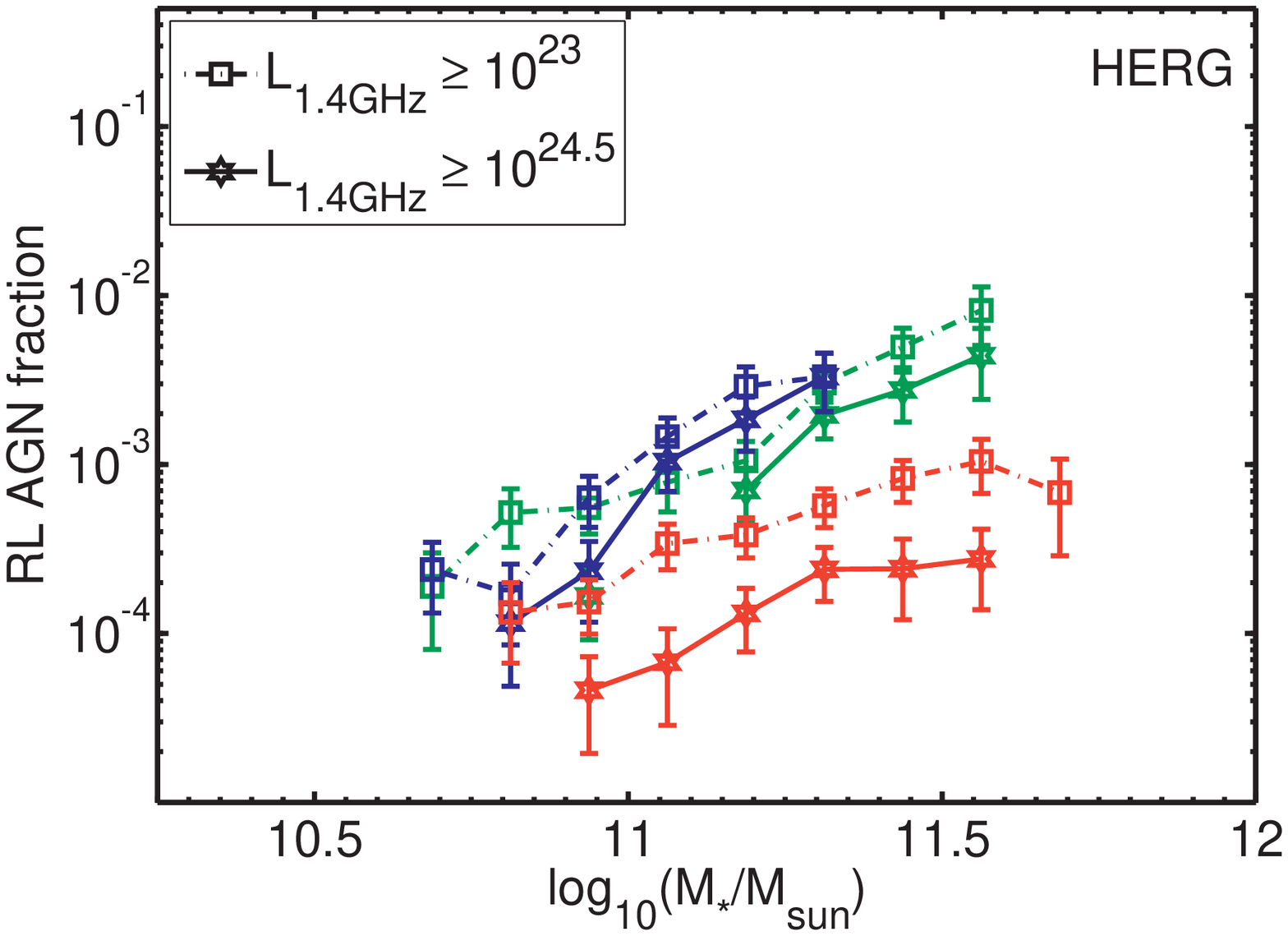}
	\end{minipage}
\caption{The fraction of red, green and blue galaxies hosting a LERG (left panel) or HERG (right panel) as a function of stellar mass. In both panels two radio luminosity cuts are shown per color. The errorbars are determined by Poissonian statistics.}
\label{FigFracSM}
\end{figure*}

\begin{figure*}
	\begin{minipage}{0.48\textwidth}
	\centering
	\includegraphics[width=1\textwidth,keepaspectratio]{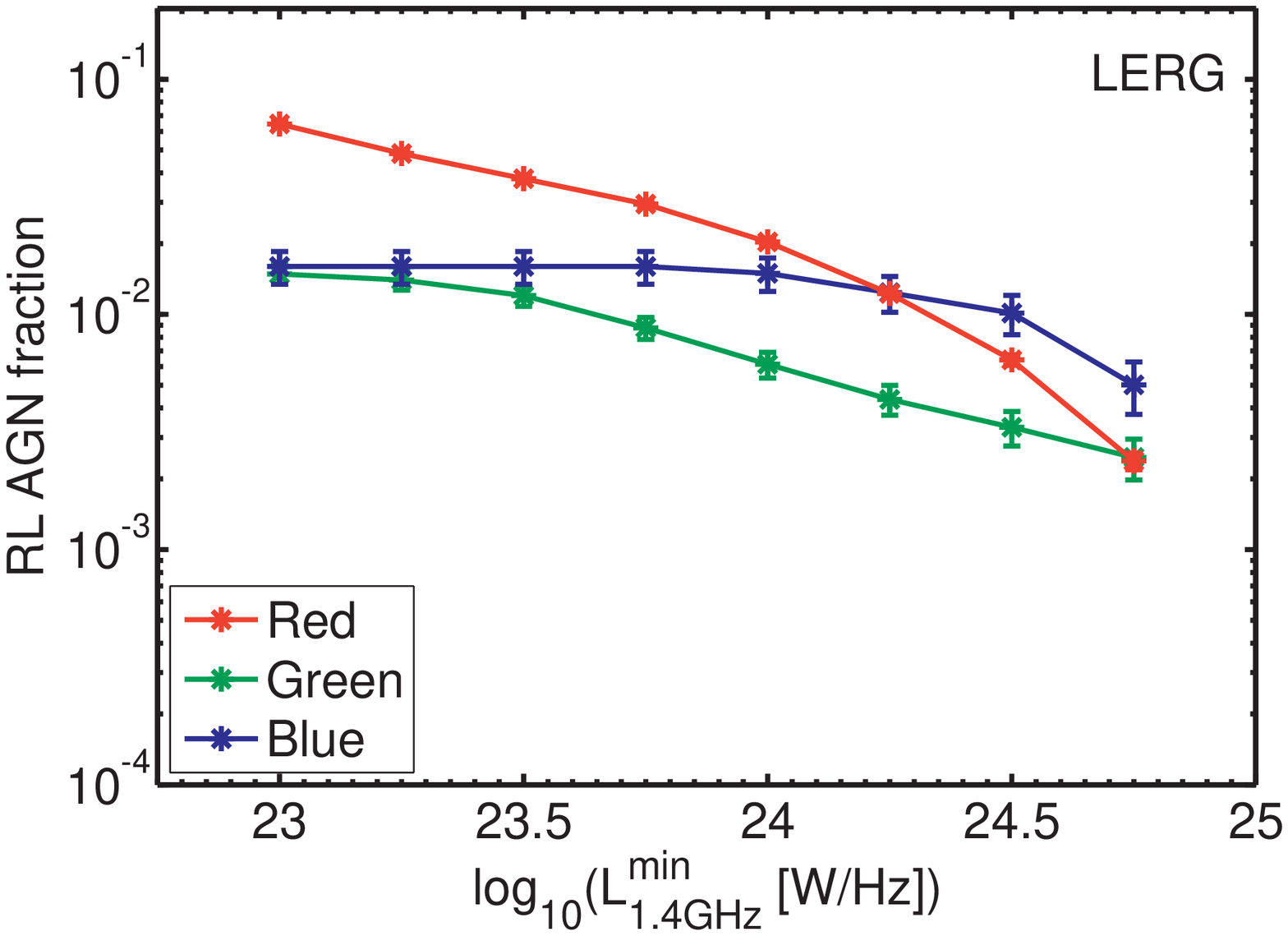}
	\end{minipage}
	\hspace{0.04\textwidth}
	\begin{minipage}{0.48\textwidth}
	\centering
	\includegraphics[width=1\textwidth,keepaspectratio]{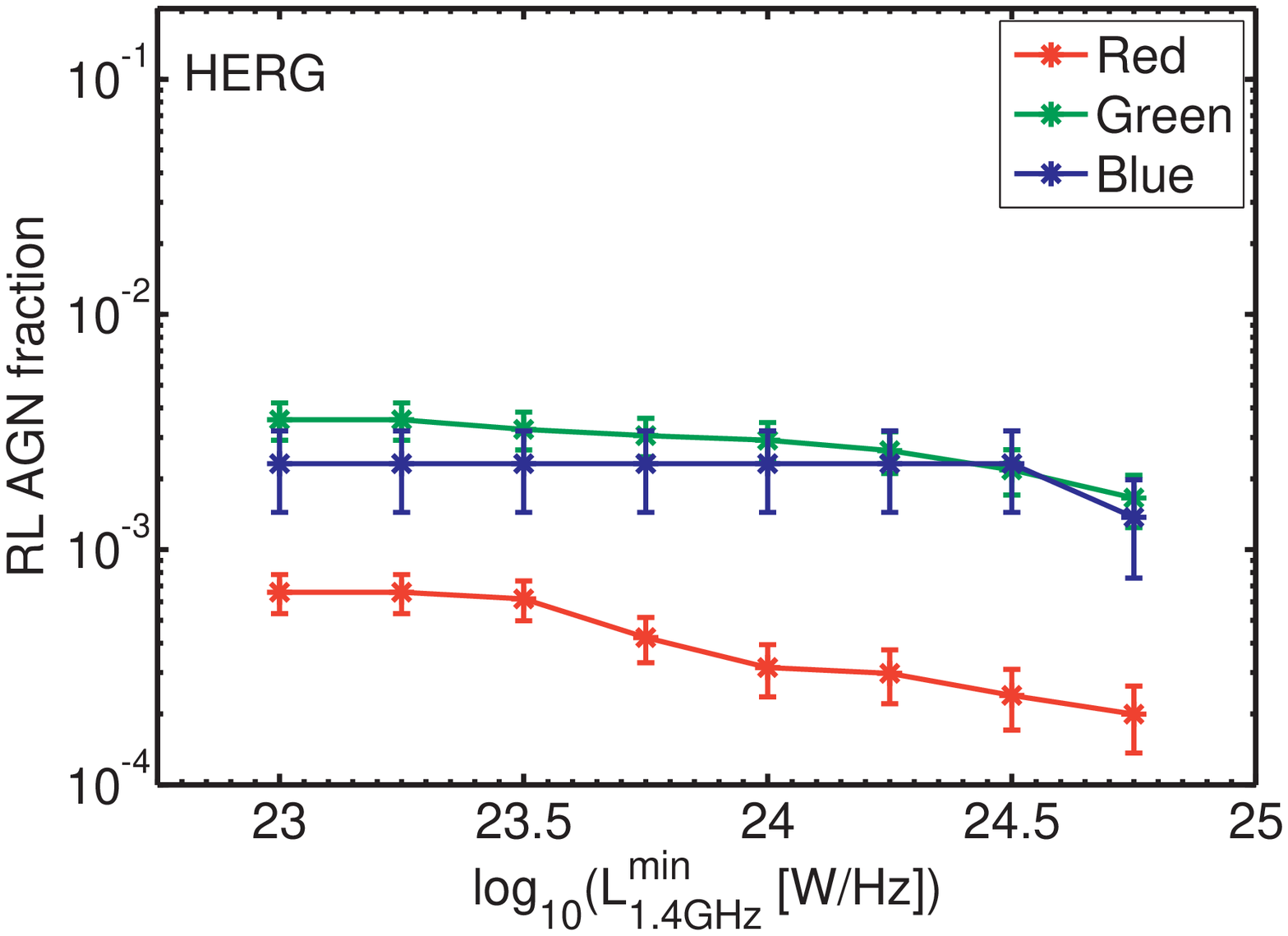}
	\end{minipage}
\caption{The fraction of red, green and blue galaxies with a mass $10^{11.25} \leq M_*/\Msun \leq 10^{11.5}$, which host a LERG (left panel) or HERG (right panel), as a function of radio luminosity cut. The errorbars are determined by Poissonian statistics.}
\label{FigFracRL}
\end{figure*}

\begin{figure*}
	\begin{minipage}{0.48\textwidth}
	\centering
	\includegraphics[width=1\textwidth,keepaspectratio]{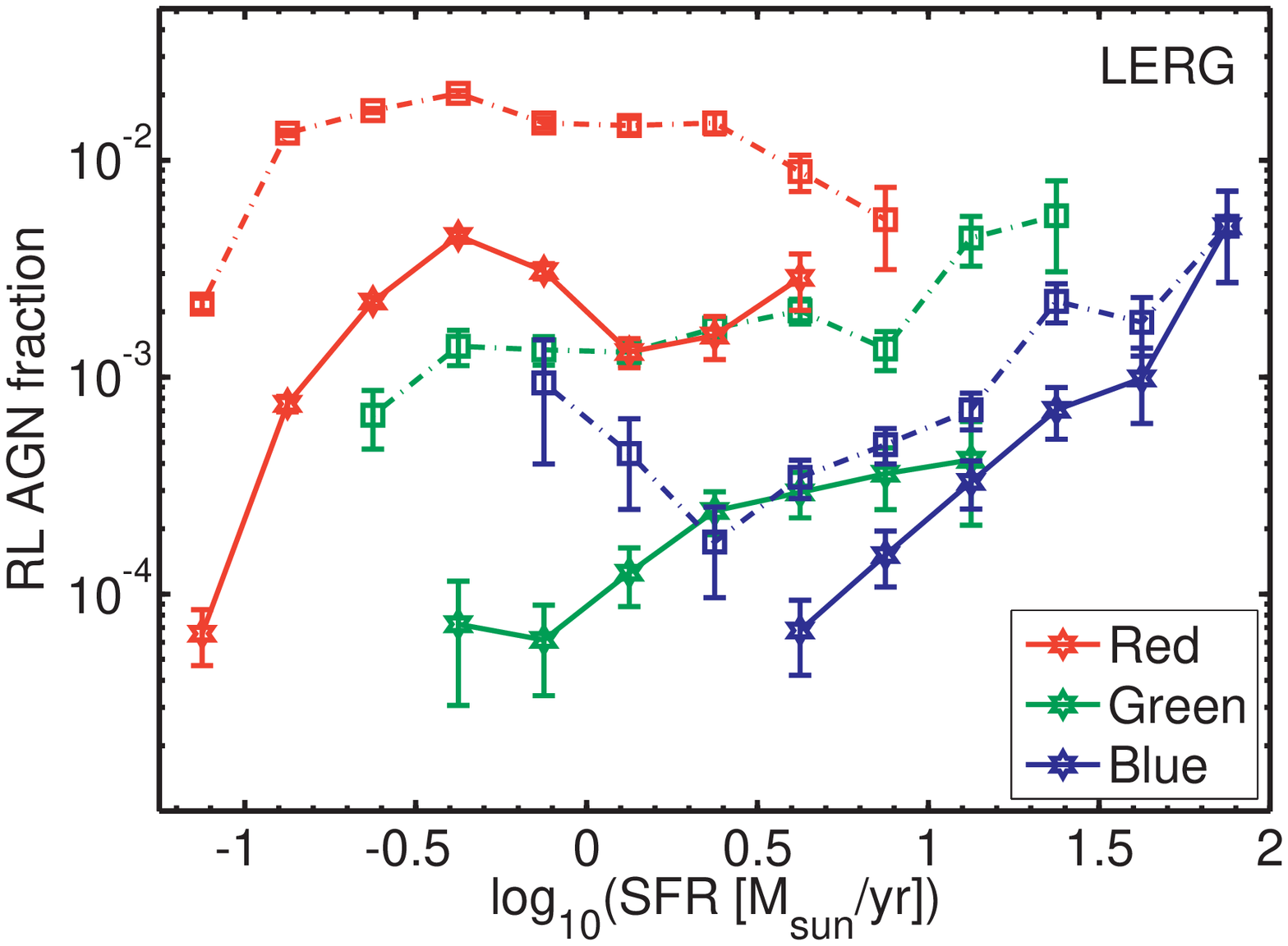}
	\end{minipage}
	\hspace{0.04\textwidth}
	\begin{minipage}{0.48\textwidth}
	\centering
	\includegraphics[width=1\textwidth,keepaspectratio]{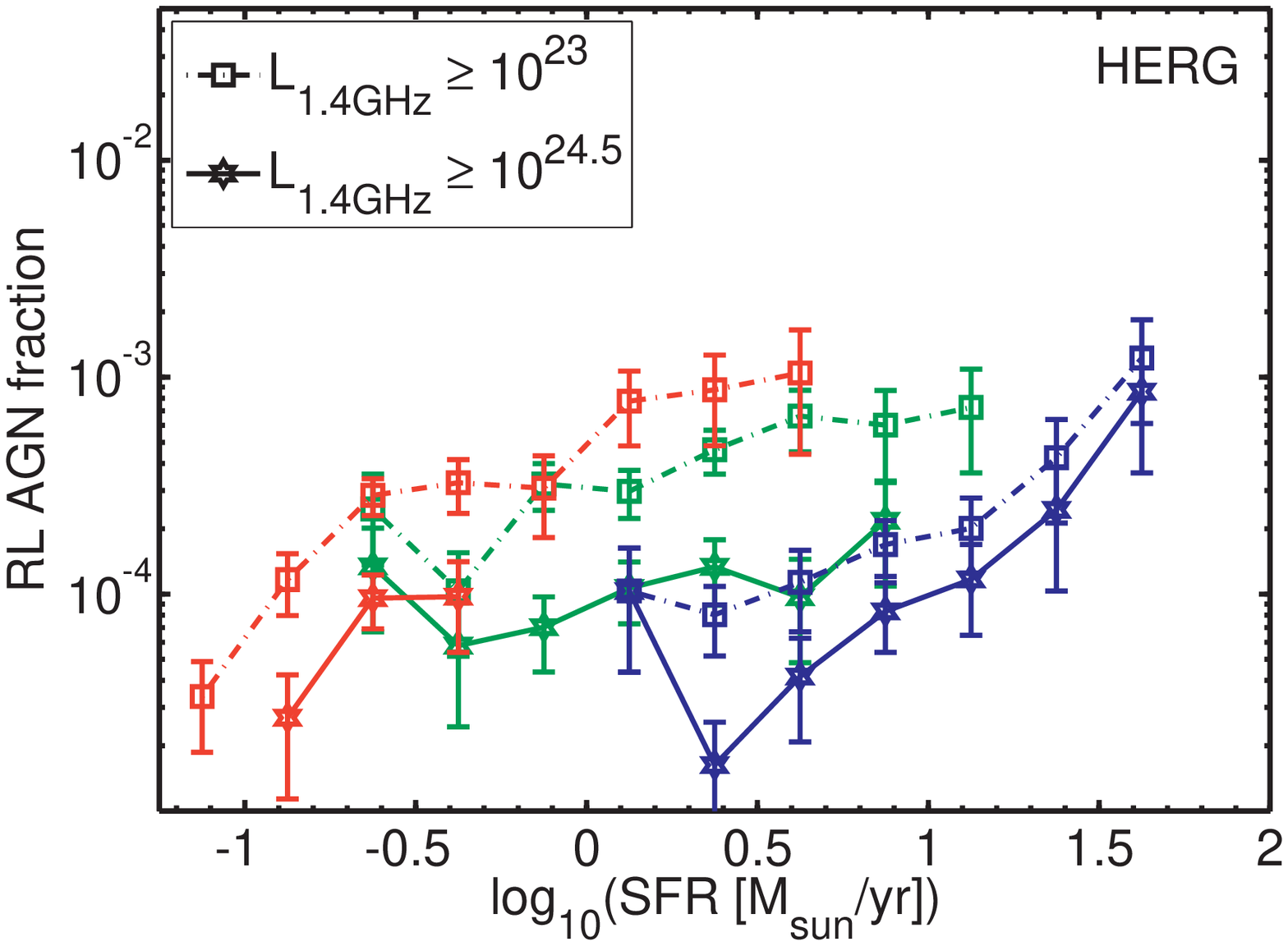}
	\end{minipage}
\caption{The fraction of red, green and blue galaxies hosting a LERG (left panel) or HERG (right panel) as a function of star formation rate. In both panels two radio luminosity cuts are shown per color. The errorbars are determined by Poissonian statistics.}
\label{FigFracSFR}
\end{figure*}

The left-hand panel of Fig. \ref{FigFracSM} shows the fraction of red, green and blue galaxies that is host to a LERG as a function of host galaxy stellar mass and radio luminosity cut. For all combinations of color and radio luminosity cut the fraction of galaxies hosting a LERG is consistent with the $f^{LE,x}_{RL} \propto M^{2.5}_*$ power law dependence typical for LERG. A saturation at high masses similar to Fig. \ref{FigFracERG} is found as well. However, the most striking feature of Fig. \ref{FigFracSM} (left panel) is that for an increasing value of the radio luminosity cut the fraction of LERGs in the red and green galaxy population drops strongly, while in the blue population it is relatively unchanged. The left-hand panel of Fig. \ref{FigFracRL} shows this effect in more detail. It shows the fraction of galaxies with a mass $10^{11.25} \leq M_*/\Msun \leq 10^{11.5}$ that host a LERG with a radio luminosity above a given $L^{\min}_{1.4GHz}$.The mass range $10^{11.25} \leq M_*/\Msun \leq 10^{11.5}$ was chosen because it has the highest number of LERGs in green (36) and blue (28) hosts. Because of their scarceness, these LERG populations limit our statistics. At low $L^{\min}_{1.4GHz}$ the probability of finding a LERG in a red galaxy is significantly greater than the probability of finding one in a blue galaxy. However, for increasing values of $L_{1.4GHz}^{\min}$, $f_{RL}^{LE,red}$ decreases, while $f_{RL}^{LE,blue}$ remains almost constant out to $L_{1.4GHz}^{\min} \approx 10^{24.5}$ $\WHz$. As a result, there is a cross-over at $L_{1.4GHz}^{\min} \approx 10^{24.25} \ \WHz$. At higher radio luminosities the RL AGN fraction can be higher for the blue galaxy population than it is for the red. For example, at a $L_{1.4GHz} \geq 10^{24.5} \ \WHz$ the fraction of RL AGN in a blue host, $f^{LE,blue}_{RL}$, with a mass in the range $10^{11.25}\ \Msun < M_* < 10^{11.5}\ \Msun$, is $1.58\pm 0.31$ times higher than $f^{LE,red}_{RL}$.\\
As can be seen in Fig. \ref{FigFracSM} and particularly in Fig. \ref{FigFracRL}, the green population has a LERG fraction, $f^{LE,green}_{RL}$, which is generally lower than $f^{LE,red}_{RL}$ and equal to or lower than $f^{LE,blue}_{RL}$. Green galaxies are less likely to host LERGs than either blue or red galaxies. In the example given above $f^{LE,green}_{RL}/f^{LE,red}_{RL}=0.52\pm 0.09$. Only at the highest radio luminosity cut considered, $L_{1.4GHz}^{\min} \geq 10^{24.75}$, is there an indication of a cross-over between $f^{LE,red}_{RL}$ and $f^{LE,green}_{RL}$. This is the result of the LERG fraction in green galaxies decreasing more slowly with increasing radio luminosity than that in red galaxies.\\
The right-hand panels of Fig. \ref{FigFracSM} and \ref{FigFracRL} are graphs equivalent to the left-hand panels, only for the HERG population. Figure \ref{FigFracSM} shows that HERG have a preference for galaxies with a bluer color. The mass dependence of $f^{HE,x}_{RL}$ seems fairly similar for all three colors. However, the probability of finding a HERG in a blue or green galaxy is significantly higher than for a red galaxy, particularly at high masses. This preference is illustrated even more clearly in Fig. \ref{FigFracRL}. For galaxies with a mass $10^{11.25}\ \Msun < M_* < 10^{11.5}\ \Msun$ green and blue galaxies are most likely to host a HERG. The probability of a red galaxy hosting a HERG is significantly lower at all radio luminosities. At $L^{min}_{1.4GHz}=10^{24} \ \WHz$ the difference in hosting probability between green and red is a factor of $f^{HE,green}_{RL}/f^{HE,red}_{RL}=9.3\pm2.9$.
It can also be clearly seen in the right-hand panel of Fig. \ref{FigFracRL}, that the fraction of galaxies hosting a HERG is almost independent of radio luminosity cut for blue and green galaxies. For the LERGs this was only observed in blue galaxies. The fraction of red galaxies hosting a HERG shows a slope, but this is significantly shallower than that of the LERG hosting fraction of red and green galaxies.\\
Figure \ref{FigFracSFR} shows the RL AGN fraction as a function of SFR and color for LERGs (left panel) and HERGs (right panel). The fraction of blue galaxies hosting a RL AGN of either excitation increases strongly with increasing SFR. A linear least squares fit of the data at a radio luminosity cut $L_{1.4GHz} \geq 10^{24} \ \WHz$ shows $f^{LE,blue}_{RL} \propto \mathrm{SFR}^{1.16 \pm 0.15}$ for LERG and $f^{HE,blue}_{RL} \propto \mathrm{SFR}^{1.24 \pm 0.27}$ for HERG. In addition, the normalization of $f^{y,blue}_{RL}$ is fairly similar for both excitations. This similarity in slope and normalization is also found in the green galaxy population. In contrast to the blue galaxy population, the fraction of green galaxies that hosts either a HERG or a LERG hardly shows any SFR dependence. The red population does show a difference in its $f^{y,red}_{RL}(\mathrm{SFR})$ when comparing HERG and LERG. The probability for a red galaxy to host a LERG is typically 1 to 1.5 orders of magnitude higher than it is to host a HERG. $f^{y,red}_{RL}$ generally does not depend strongly on SFR. However, if a trend had to be identified this would be increasing with SFR for the red HERG population and decreasing with SFR for the red LERG population. All these trends shown in Fig. \ref{FigFracSFR} are largely independent of the radio luminosity cut.\\
\section{\label{Discussion}Discussion}
The results presented in the previous section generally support the ``radio mode'' vs ``quasar mode'' AGN paradigm in which RL AGN are preferentially hosted by old (red) elliptical galaxies and radio-loud quasars by younger, star forming systems \citep{Croton2006,Hardcastle2007}. It is thought that the former are powered through the accretion of hot halo gas. Our results support this scenario, as LERGs of all colors show the $f^{LE,x}_{RL} \propto M^{2.5}_*$ power law, which is indicative of the accretion of hot halo gas \citep{Best2006}. However, additional processes must be occurring when cold gas is present inside the host galaxy. Our results indicate that the probability that one of the rare massive ($M_* > 10^{10.8} \Msun$) blue galaxies is host to a luminous $(L_{1.4GHz} \gtrsim 10^{24.5} \ \WHz)$ LERG is higher than for one of the many red galaxies with a similar mass. It is the scarceness of massive blue galaxies that results in the fact that red LERGs dominate the radio AGN population in the local Universe; this is a dominance, which can already be seen from the absolute numbers in Table \ref{GalaxyNumbers}. Red LERGs are ten times more numerous than either LERGs of other colors or HERGs, even at moderate radio luminosities ($L_{1.4GHz} \geq 10^{24.5} \ \WHz$). In addition, the absolute value of $f^{LE,red}_{RL}$ is generally higher than for other combination of $(y,x)$.\\
The strong SFR dependence of $f^{LE,blue}_{RL}$ found for the massive blue galaxies corroborates the idea that the enhanced probability and radio luminosity of blue LERG is the result of the presence of cold gas within the host galaxy. We speculate that this cold gas is the result of the higher cooling rates required to support the increased accretion rate needed for a more powerful AGN. This increased amount of cooling gas could result in a level of cold gas within the host that equates to a significant amount of star formation. Alternatively, the cold gas could be brought into the host galaxy through merging events or tidal interactions with nearby galaxies. This supply mechanism for cold gas is supported by a visual inspection of the SDSS images of all red, green and blue radio galaxies with a redshift $z \leq 0.1$. At least five of the 18 blue LERGs show morphological signs (such as tails) indicative of a merging event. Around 5\% of the red LERGs show signs of a merging event as well. However, all these mergers are in a very early stage. This could be concluded from the fact that the two original nuclei could still be identified in all cases.\\ 
The most important result with respect to the HERGs is the completely different mass dependence compared to LERGs. The fraction of galaxies hosting a HERG follows a $f^{HE}_{RL} \propto M^{1.5}_*$ power law. This puts them, as expected, more in line with optically identified (``quasar-mode'') AGN \citep{Best2005b}. HERGs are thought to be accreting cold gas supplied through galaxy interactions, which is in line with the fact that $\sim 25\%$ of the HERGs at $z<0.1$ show signs of a past merging event in their morphology. Galaxy interaction are less dependent on galaxy mass, hence the lower mass dependence of the hosting probability. However, the fact that HERGs still have some mass dependence and optically identified AGN do not, suggests a mass dependence of radio-loudness even in quasar-like states. In addition, it is worth noting that at the lowest masses considered, $10^{10.25} \leq M_*/\Msun \leq 10^{10.5}$, the fraction of galaxies hosting a HERG and LERG is approximately equal. This is the mass range in which \citet{Tasse2008b} found a switch between HERG and LERG domination.\\
Green galaxies are generally considered to be the transitional stage when galaxies evolve from the blue cloud to the red sequence. Therefore their properties are generally expected to be at intermediate levels, smoothly connecting the properties of the blue and red galaxy population \citep{Martin2007}. In this work we find this is not true with respect to green galaxies hosting radio galaxies of either excitation. The left-hand panels of figures \ref{FigFracSM} and \ref{FigFracRL} show that the probability of hosting a LERG of a fixed radio luminosity is lower for a green galaxy than for a red or blue galaxy of similar mass. In addition, the probability of finding a HERG in a given galaxy peaks within the green population. These results are also a clear indication that the green galaxy sample identified here is not just a mixture of poorly separated blue and red galaxies. In that case all trends would be in between those of red and blue, which is not the case. We speculate that the reduced probability of finding a LERG in a green galaxy is due AGN feedback \citep{Croton2006}. The quasar winds and radio jets that are the result of the high-excitation AGN activity in massive blue and green galaxies will blow the remaining cold gas out of the galaxy, stopping star formation and cold gas accretion. Meanwhile, the radio jets also reheat the gas halo, which temporarily reduces the cooling flow of the halo gas to the SMBH. This effectively stops all AGN activity and allows the stellar population to age across the green valley. Only after the galaxy has reached the red sequence will the halo gas have cooled enough to re-establish the cooling flow and trigger a new episode of RL AGN activity through hot gas accretion.\\
\section{\label{Conclusions}Conclusions}
We investigated the fraction of massive blue, green and red galaxies that are host to a RL AGN. In this we consider RL AGN that are classified as LERG separately from those classified as HERG. It is shown that:
\begin{enumerate}
\item RL AGN in the local Universe are dominated by LERGs. This is true even if the red, green and blue galaxy population are considered individually. The probability that a galaxy of any color hosts a LERG follows a $f^{LE,x}_{RL} \propto M^{2.5}_*$ power law. This is much steeper than what is found for HERGs: $f^{HE}_{RL} \propto M^{1.5}_*$. 
\item The probability that a red galaxy of a given mass hosts a LERG depends strongly on the applied radio luminosity cut. This radio power dependence is much weaker for HERG of any color and LERG hosted by the blue galaxy population.
\item The probability that a green galaxy of a given mass is host to a LERG is lower compared to both the red and the blue galaxy population by a factor of 2-3 for $L_{1.4GHz}^{\min} \approx 10^{24.5} \ \WHz$ . The $f_{RL}^{LE,green}$ depends on $L_{1.4GHz}^{\min}$ in a similar way as $f_{RL}^{LE,red}$, but the slope is slightly weaker. However, the probability to host a HERG in a blue or green galaxy is significantly higher than it is in red galaxies. The maximum hosting probability for HERG is found for $M_* \approx 10^{11.5} \ \Msun$ green galaxies with a typical $\mathrm{SFR} = 3 \ \Myr$. The hosting probability of HERG deceases towards lower masses and SFR and the maximum shifts to bluer galaxies for more luminous radio sources.
\item Within the blue galaxy population the probability to host a RL AGN (both LERG and HERG) shows a $f^{y,blue}_{RL} \propto \mathrm{SFR}$ dependence. This SFR dependence is weaker or absent in the red and green population. However, the absolute value of the hosting probability of blue galaxies is too low to have an effect on the AGN population as a whole.
\end{enumerate}
These results suggest that the presence of cold gas in a galaxy enhances the probability that its SMBH becomes a luminous $(L_{1.4GHz} \geq 10^{24.5} \ \WHz)$ RL AGN compared to the typical ``model'' LERG in red elliptical galaxies. However, the presence of cold gas with the host galaxy does not automatically generate a high-excitation AGN. HERGs clearly prefer galaxies rich in cold gas and typically host luminous AGN. We speculate that feedback of this AGN activity is responsible for the reduced probability of green galaxies to host a LERG.
\begin{acknowledgements}
R.M.J. Janssen would like to thank A. Endo and T.M. Klapwijk for facilitating this research and NOVA for the financial support. P.N. Best is grateful for financial support from the Leverhulme Trust.\\
The research makes use of the NVSS and FIRST radio surveys and SDSS optical survey. The NVSS and FIRST radio surveys were carried out using the National Radio Astronomy Observatory Very Large Array: NRAO is operated by Associated Universities Inc., under co-operative agreement with the National Science Foundation.\\
Funding for the SDSS and SDSS-II has been provided by the Alfred P. Sloan Foundation, the Participating Institutions, the National Science Foundation, the U.S. Department of Energy, the National Aeronautics and Space Administration, the Japanese Monbukagakusho, the Max Planck Society, and the Higher Education Funding Council for England. The SDSS Web Site is http://www.sdss.org/.\\
The SDSS is managed by the Astrophysical Research Consortium for the Participating Institutions. The Participating Institutions are the American Museum of Natural History, Astrophysical Institute Potsdam, University of Basel, University of Cambridge, Case Western Reserve University, University of Chicago, Drexel University, Fermilab, the Institute for Advanced Study, the Japan Participation Group, Johns Hopkins University, the Joint Institute for Nuclear Astrophysics, the Kavli Institute for Particle Astrophysics and Cosmology, the Korean Scientist Group, the Chinese Academy of Sciences (LAMOST), Los Alamos National Laboratory, the Max-Planck-Institute for Astronomy (MPIA), the Max-Planck-Institute for Astrophysics (MPA), New Mexico State University, Ohio State University, University of Pittsburgh, University of Portsmouth, Princeton University, the United States Naval Observatory, and the University of Washington.\\
\end{acknowledgements}

\bibliographystyle{aa}
\bibliography{BiblioAGN2012}

\begin{thebibliography}{29}
\expandafter\ifx\csname natexlab\endcsname\relax\def\natexlab#1{#1}\fi

\bibitem[{{Abazajian} {et~al.}(2009){Abazajian}, {Adelman-McCarthy},
  {Ag{\"u}eros}, {Allam}, {Allende Prieto}, {An}, {Anderson}, {Anderson},
  {Annis}, {Bahcall}, \& et~al.}]{Abazajian2009}
{Abazajian}, K.~N., {Adelman-McCarthy}, J.~K., {Ag{\"u}eros}, M.~A., {et~al.}
  2009, \apjs, 182, 543

\bibitem[{{Baldwin} {et~al.}(1981){Baldwin}, {Phillips}, \&
  {Terlevich}}]{Baldwin1981}
{Baldwin}, J.~A., {Phillips}, M.~M., \& {Terlevich}, R. 1981, \pasp, 93, 5

\bibitem[{{Balogh} {et~al.}(1999){Balogh}, {Morris}, {Yee}, {Carlberg}, \&
  {Ellingson}}]{Balogh1999}
{Balogh}, M.~L., {Morris}, S.~L., {Yee}, H.~K.~C., {Carlberg}, R.~G., \&
  {Ellingson}, E. 1999, \apj, 527, 54

\bibitem[{{Becker} {et~al.}(1995){Becker}, {White}, \& {Helfand}}]{Becker1995}
{Becker}, R.~H., {White}, R.~L., \& {Helfand}, D.~J. 1995, \apj, 450, 559

\bibitem[{{Best} \& {Heckman}(2012)}]{Best2012}
{Best}, P.~N. \& {Heckman}, T.~M. 2012, \mnras

\bibitem[{{Best} {et~al.}(2006){Best}, {Kaiser}, {Heckman}, \&
  {Kauffmann}}]{Best2006}
{Best}, P.~N., {Kaiser}, C.~R., {Heckman}, T.~M., \& {Kauffmann}, G. 2006,
  \mnras, 368, L67

\bibitem[{{Best} {et~al.}(2005{\natexlab{a}}){Best}, {Kauffmann}, {Heckman},
  {Brinchmann}, {Charlot}, {Ivezi{\'c}}, \& {White}}]{Best2005b}
{Best}, P.~N., {Kauffmann}, G., {Heckman}, T.~M., {et~al.} 2005{\natexlab{a}},
  \mnras, 362, 25

\bibitem[{{Best} {et~al.}(2005{\natexlab{b}}){Best}, {Kauffmann}, {Heckman}, \&
  {Ivezi{\'c}}}]{Best2005a}
{Best}, P.~N., {Kauffmann}, G., {Heckman}, T.~M., \& {Ivezi{\'c}}, {\v Z}.
  2005{\natexlab{b}}, \mnras, 362, 9

\bibitem[{{Brinchmann} {et~al.}(2004){Brinchmann}, {Charlot}, {White},
  {Tremonti}, {Kauffmann}, {Heckman}, \& {Brinkmann}}]{Brinchmann2004b}
{Brinchmann}, J., {Charlot}, S., {White}, S.~D.~M., {et~al.} 2004, \mnras, 351,
  1151

\bibitem[{{Condon} {et~al.}(1998){Condon}, {Cotton}, {Greisen}, {Yin},
  {Perley}, {Taylor}, \& {Broderick}}]{Condon1998}
{Condon}, J.~J., {Cotton}, W.~D., {Greisen}, E.~W., {et~al.} 1998, \aj, 115,
  1693

\bibitem[{{Croton} {et~al.}(2006){Croton}, {Springel}, {White}, {De Lucia},
  {Frenk}, {Gao}, {Jenkins}, {Kauffmann}, {Navarro}, \& {Yoshida}}]{Croton2006}
{Croton}, D.~J., {Springel}, V., {White}, S.~D.~M., {et~al.} 2006, \mnras, 365,
  11

\bibitem[{{Evans} {et~al.}(2006){Evans}, {Worrall}, {Hardcastle}, {Kraft}, \&
  {Birkinshaw}}]{Evans2006}
{Evans}, D.~A., {Worrall}, D.~M., {Hardcastle}, M.~J., {Kraft}, R.~P., \&
  {Birkinshaw}, M. 2006, \apj, 642, 96

\bibitem[{{Hardcastle} {et~al.}(2007){Hardcastle}, {Evans}, \&
  {Croston}}]{Hardcastle2007}
{Hardcastle}, M.~J., {Evans}, D.~A., \& {Croston}, J.~H. 2007, \mnras, 376,
  1849

\bibitem[{{Hickox} {et~al.}(2009){Hickox}, {Jones}, {Forman}, {Murray},
  {Kochanek}, {Eisenstein}, {Jannuzi}, {Dey}, {Brown}, {Stern}, {Eisenhardt},
  {Gorjian}, {Brodwin}, {Narayan}, {Cool}, {Kenter}, {Caldwell}, \&
  {Anderson}}]{Hickox2009}
{Hickox}, R.~C., {Jones}, C., {Forman}, W.~R., {et~al.} 2009, \apj, 696, 891

\bibitem[{{Kauffmann} {et~al.}(2008){Kauffmann}, {Heckman}, \&
  {Best}}]{Kauffmann2008}
{Kauffmann}, G., {Heckman}, T.~M., \& {Best}, P.~N. 2008, \mnras, 384, 953

\bibitem[{{Kauffmann} {et~al.}(2003{\natexlab{a}}){Kauffmann}, {Heckman},
  {White}, {Charlot}, {Tremonti}, {Brinchmann}, {Bruzual}, {Peng}, {Seibert},
  {Bernardi}, {Blanton}, {Brinkmann}, {Castander}, {Cs{\'a}bai}, {Fukugita},
  {Ivezic}, {Munn}, {Nichol}, {Padmanabhan}, {Thakar}, {Weinberg}, \&
  {York}}]{Kauffmann2003a}
{Kauffmann}, G., {Heckman}, T.~M., {White}, S.~D.~M., {et~al.}
  2003{\natexlab{a}}, \mnras, 341, 33

\bibitem[{{Kauffmann} {et~al.}(2003{\natexlab{b}}){Kauffmann}, {Heckman},
  {White}, {Charlot}, {Tremonti}, {Peng}, {Seibert}, {Brinkmann}, {Nichol},
  {SubbaRao}, \& {York}}]{Kauffmann2003b}
{Kauffmann}, G., {Heckman}, T.~M., {White}, S.~D.~M., {et~al.}
  2003{\natexlab{b}}, \mnras, 341, 54

\bibitem[{{Laing} {et~al.}(1994){Laing}, {Jenkins}, {Wall}, \&
  {Unger}}]{Laing1994}
{Laing}, R.~A., {Jenkins}, C.~R., {Wall}, J.~V., \& {Unger}, S.~W. 1994, in
  Astronomical Society of the Pacific Conference Series, Vol.~54, The Physics
  of Active Galaxies, ed. {G.~V.~Bicknell, M.~A.~Dopita, \& P.~J.~Quinn}, 201

\bibitem[{{Martin} {et~al.}(2007){Martin}, {Wyder}, {Schiminovich}, {Barlow},
  {Forster}, {Friedman}, {Morrissey}, {Neff}, {Seibert}, {Small}, {Welsh},
  {Bianchi}, {Donas}, {Heckman}, {Lee}, {Madore}, {Milliard}, {Rich}, {Szalay},
  \& {Yi}}]{Martin2007}
{Martin}, D.~C., {Wyder}, T.~K., {Schiminovich}, D., {et~al.} 2007, \apjs, 173,
  342

\bibitem[{{Matthews} {et~al.}(1964){Matthews}, {Morgan}, \&
  {Schmidt}}]{Matthews1964}
{Matthews}, T.~A., {Morgan}, W.~W., \& {Schmidt}, M. 1964, \apj, 140, 35

\bibitem[{{Schmidt}(1968)}]{Schmidt1968}
{Schmidt}, M. 1968, \apj, 151, 393

\bibitem[{{Stoughton} {et~al.}(2002){Stoughton}, {Lupton}, {Bernardi},
  {Blanton}, {Burles}, {Castander}, {Connolly}, {Eisenstein}, {Frieman},
  {Hennessy}, {Hindsley}, {Ivezi{\'c}}, {Kent}, {Kunszt}, {Lee}, {Meiksin},
  {Munn}, {Newberg}, {Nichol}, {Nicinski}, {Pier}, {Richards}, {Richmond},
  {Schlegel}, {Smith}, {Strauss}, {SubbaRao}, {Szalay}, {Thakar}, {Tucker},
  {Vanden Berk}, {Yanny}, {Adelman}, {Anderson}, {Anderson}, {Annis},
  {Bahcall}, {Bakken}, {Bartelmann}, {Bastian}, {Bauer}, {Berman},
  {B{\"o}hringer}, {Boroski}, {Bracker}, {Briegel}, {Briggs}, {Brinkmann},
  {Brunner}, {Carey}, {Carr}, {Chen}, {Christian}, {Colestock}, {Crocker},
  {Csabai}, {Czarapata}, {Dalcanton}, {Davidsen}, {Davis}, {Dehnen},
  {Dodelson}, {Doi}, {Dombeck}, {Donahue}, {Ellman}, {Elms}, {Evans}, {Eyer},
  {Fan}, {Federwitz}, {Friedman}, {Fukugita}, {Gal}, {Gillespie}, {Glazebrook},
  {Gray}, {Grebel}, {Greenawalt}, {Greene}, {Gunn}, {de Haas}, {Haiman},
  {Haldeman}, {Hall}, {Hamabe}, {Hansen}, {Harris}, {Harris}, {Harvanek},
  {Hawley}, {Hayes}, {Heckman}, {Helmi}, {Henden}, {Hogan}, {Hogg}, {Holmgren},
  {Holtzman}, {Huang}, {Hull}, {Ichikawa}, {Ichikawa}, {Johnston}, {Kauffmann},
  {Kim}, {Kimball}, {Kinney}, {Klaene}, {Kleinman}, {Klypin}, {Knapp},
  {Korienek}, {Krolik}, {Kron}, {Krzesi{\'n}ski}, {Lamb}, {Leger},
  {Limmongkol}, {Lindenmeyer}, {Long}, {Loomis}, {Loveday}, {MacKinnon},
  {Mannery}, {Mantsch}, {Margon}, {McGehee}, {McKay}, {McLean}, {Menou},
  {Merelli}, {Mo}, {Monet}, {Nakamura}, {Narayanan}, {Nash}, {Neilsen},
  {Newman}, {Nitta}, {Odenkirchen}, {Okada}, {Okamura}, {Ostriker}, {Owen},
  {Pauls}, {Peoples}, {Peterson}, {Petravick}, {Pope}, {Pordes}, {Postman},
  {Prosapio}, {Quinn}, {Rechenmacher}, {Rivetta}, {Rix}, {Rockosi}, {Rosner},
  {Ruthmansdorfer}, {Sandford}, {Schneider}, {Scranton}, {Sekiguchi}, {Sergey},
  {Sheth}, {Shimasaku}, {Smee}, {Snedden}, {Stebbins}, {Stubbs}, {Szapudi},
  {Szkody}, {Szokoly}, {Tabachnik}, {Tsvetanov}, {Uomoto}, {Vogeley}, {Voges},
  {Waddell}, {Walterbos}, {Wang}, {Watanabe}, {Weinberg}, {White}, {White},
  {Wilhite}, {Wolfe}, {Yasuda}, {York}, {Zehavi}, \& {Zheng}}]{Stoughton2002}
{Stoughton}, C., {Lupton}, R.~H., {Bernardi}, M., {et~al.} 2002, \aj, 123, 485

\bibitem[{{Strateva} {et~al.}(2001){Strateva}, {Ivezi{\'c}}, {Knapp},
  {Narayanan}, {Strauss}, {Gunn}, {Lupton}, {Schlegel}, {Bahcall}, {Brinkmann},
  {Brunner}, {Budav{\'a}ri}, {Csabai}, {Castander}, {Doi}, {Fukugita}, {Gy{\H
  o}ry}, {Hamabe}, {Hennessy}, {Ichikawa}, {Kunszt}, {Lamb}, {McKay},
  {Okamura}, {Racusin}, {Sekiguchi}, {Schneider}, {Shimasaku}, \&
  {York}}]{Strateva2001}
{Strateva}, I., {Ivezi{\'c}}, {\v Z}., {Knapp}, G.~R., {et~al.} 2001, \aj, 122,
  1861

\bibitem[{{Strauss} {et~al.}(2002){Strauss}, {Weinberg}, {Lupton}, {Narayanan},
  {Annis}, {Bernardi}, {Blanton}, {Burles}, {Connolly}, {Dalcanton}, {Doi},
  {Eisenstein}, {Frieman}, {Fukugita}, {Gunn}, {Ivezi{\'c}}, {Kent}, {Kim},
  {Knapp}, {Kron}, {Munn}, {Newberg}, {Nichol}, {Okamura}, {Quinn}, {Richmond},
  {Schlegel}, {Shimasaku}, {SubbaRao}, {Szalay}, {Vanden Berk}, {Vogeley},
  {Yanny}, {Yasuda}, {York}, \& {Zehavi}}]{Strauss2002}
{Strauss}, M.~A., {Weinberg}, D.~H., {Lupton}, R.~H., {et~al.} 2002, \aj, 124,
  1810

\bibitem[{{Tadhunter} {et~al.}(2005){Tadhunter}, {Robinson}, {Gonz{\'a}lez
  Delgado}, {Wills}, \& {Morganti}}]{Tadhunter2005}
{Tadhunter}, C., {Robinson}, T.~G., {Gonz{\'a}lez Delgado}, R.~M., {Wills}, K.,
  \& {Morganti}, R. 2005, \mnras, 356, 480

\bibitem[{{Tasse} {et~al.}(2008){Tasse}, {Best}, {R{\"o}ttgering}, \& {Le
  Borgne}}]{Tasse2008b}
{Tasse}, C., {Best}, P.~N., {R{\"o}ttgering}, H., \& {Le Borgne}, D. 2008,
  \aap, 490, 893

\bibitem[{{Wall} \& {Jenkins}(2003)}]{Wall2003}
{Wall}, J.~V. \& {Jenkins}, C.~R. 2003, {Practical Statistics for Astronomers},
  ed. {Wall, J.~V.~\& Jenkins, C.~R.}

\bibitem[{{Whysong} \& {Antonucci}(2004)}]{Whysong2004}
{Whysong}, D. \& {Antonucci}, R. 2004, \apj, 602, 116

\bibitem[{{York} {et~al.}(2000){York}, {Adelman}, {Anderson}, {Anderson},
  {Annis}, {Bahcall}, {Bakken}, {Barkhouser}, {Bastian}, {Berman}, {Boroski},
  {Bracker}, {Briegel}, {Briggs}, {Brinkmann}, {Brunner}, {Burles}, {Carey},
  {Carr}, {Castander}, {Chen}, {Colestock}, {Connolly}, {Crocker}, {Csabai},
  {Czarapata}, {Davis}, {Doi}, {Dombeck}, {Eisenstein}, {Ellman}, {Elms},
  {Evans}, {Fan}, {Federwitz}, {Fiscelli}, {Friedman}, {Frieman}, {Fukugita},
  {Gillespie}, {Gunn}, {Gurbani}, {de Haas}, {Haldeman}, {Harris}, {Hayes},
  {Heckman}, {Hennessy}, {Hindsley}, {Holm}, {Holmgren}, {Huang}, {Hull},
  {Husby}, {Ichikawa}, {Ichikawa}, {Ivezi{\'c}}, {Kent}, {Kim}, {Kinney},
  {Klaene}, {Kleinman}, {Kleinman}, {Knapp}, {Korienek}, {Kron}, {Kunszt},
  {Lamb}, {Lee}, {Leger}, {Limmongkol}, {Lindenmeyer}, {Long}, {Loomis},
  {Loveday}, {Lucinio}, {Lupton}, {MacKinnon}, {Mannery}, {Mantsch}, {Margon},
  {McGehee}, {McKay}, {Meiksin}, {Merelli}, {Monet}, {Munn}, {Narayanan},
  {Nash}, {Neilsen}, {Neswold}, {Newberg}, {Nichol}, {Nicinski}, {Nonino},
  {Okada}, {Okamura}, {Ostriker}, {Owen}, {Pauls}, {Peoples}, {Peterson},
  {Petravick}, {Pier}, {Pope}, {Pordes}, {Prosapio}, {Rechenmacher}, {Quinn},
  {Richards}, {Richmond}, {Rivetta}, {Rockosi}, {Ruthmansdorfer}, {Sandford},
  {Schlegel}, {Schneider}, {Sekiguchi}, {Sergey}, {Shimasaku}, {Siegmund},
  {Smee}, {Smith}, {Snedden}, {Stone}, {Stoughton}, {Strauss}, {Stubbs},
  {SubbaRao}, {Szalay}, {Szapudi}, {Szokoly}, {Thakar}, {Tremonti}, {Tucker},
  {Uomoto}, {Vanden Berk}, {Vogeley}, {Waddell}, {Wang}, {Watanabe},
  {Weinberg}, {Yanny}, {Yasuda}, \& {SDSS Collaboration}}]{York2000}
{York}, D.~G., {Adelman}, J., {Anderson}, Jr., J.~E., {et~al.} 2000, \aj, 120,
  1579

\end{thebibliography}

\end{document}